\documentclass[aps,twocolumn,prd,preprintnumbers,amsmath,amssymb,superscriptaddress,nofootinbib]{revtex4-1}  
\usepackage[linktocpage,pagebackref=false,hidelinks]{hyperref}

\usepackage{amsmath,setspace,amsfonts,latexsym}
\usepackage{amssymb}
\usepackage{color}
\usepackage{epsfig}
\usepackage{graphicx}
\usepackage{slashed}
\usepackage{ulem}
\usepackage[export]{adjustbox}

\definecolor{White}{rgb}{1,1,1}
\definecolor{Red}{rgb}{1,0.1,0}
\definecolor{LightYellow}{rgb}{1,1,.875}
\definecolor{SteelBlue}{rgb}{.273,.508,.703}
\definecolor{navy}{rgb}{0,0,.5}
\definecolor{LightCyan}{rgb}{.875,1,1}
\definecolor{DarkRed}{rgb}{.543,0,0}
\definecolor{HotPink}{rgb}{1,.41,.70}
\definecolor{ForestGreen}{rgb}{.13,.54,.13}
\definecolor{OliveDrab}{rgb}{.42,.55,.14}
\definecolor{MediumBlue}{rgb}{0,0,.80}
\definecolor{RoyalBlue}{rgb}{.25,.41,.88}
\definecolor{DeepSkyBlue}{rgb}{0,.746,1}
\definecolor{Brown}{rgb}{0.545,0.271,0.074}
\definecolor{Purple}{rgb}{0.637,0.285,0.641}

\def\bea{\begin{eqnarray}}
\def\eea{\end{eqnarray}}
\def\bec{\begin{center}}
\def\ec{\end{center}}

\def\beq{\begin{equation}}
\def\eeq{\end{equation}}

\newcommand\lsim{\mathrel{\rlap{\lower4pt\hbox{\hskip1pt$\sim$}}
    \raise1pt\hbox{$<$}}}
\newcommand\gsim{\mathrel{\rlap{\lower4pt\hbox{\hskip1pt$\sim$}}
    \raise1pt\hbox{$>$}}}
\def\bea{\begin{eqnarray}}
\def\eea{\end{eqnarray}}
\def\ba{\begin{array}}
\def\ea{\end{array}}
\def\bc{\begin{center}}
\def\ec{\end{center}}
\def\nn{\nonumber}

\def\l{\lambda}

\def\beq{\begin{equation}}
\def\eeq{\end{equation}}

\newcommand{\GeV}{\mathrm{GeV}}

\newcommand{\TeV}{\mathrm{TeV}}

\begin{document} 

\title{Spontaneous Twin Symmetry Breaking}

\author{Tae Hyun Jung} 
\email{thjung0720@ibs.re.kr}
\affiliation{Center for Theoretical Physics of the Universe, Institute for Basic Science (IBS), Daejeon, 34051, Korea}
\affiliation{Department of Physics, Florida State University, Tallahassee, FL 32306, USA}
\date{\today}
\preprint{CTPU-19-07} 

\begin{abstract}
We consider a twin Higgs model where the $Z_2$ twin symmetry is broken spontaneously, not explicitly.
We show that introducing an exact copy of the standard model with a renormalizable Higgs portal potential is enough to realize such a scenario.
In this scenario, the $SU(4)$ breaking scale $f$ is determined by the scale where the Higgs self quartic coupling flips its sign, whose standard model prediction is ${\cal O}(10^{10}\,{\rm GeV})$.
For the misalignment of nonzero vacuum expectation values of the twin Higgs fields, it is explicitly shown that parameter tuning of ${\cal O}(m_h^2/f^2)$ is required, so we conclude that minimal setup does not solve the hierarchy problem.
We point out that the tuning can be significantly reduced ($f\sim 2.7\,{\rm TeV}$) if there are twin vectorlike leptons with a large Yukawa coupling.
\end{abstract}

\maketitle

\section{Introduction}

The approach of the twin Higgs scenarios to the little hierarchy problem is a realization of the Higgs boson as a pseudo-Goldstone boson~\cite{Chacko:2005pe}.
With introducing the twin/mirror sector of the standard model (SM),
 twin Higgs fields $H_A$ and $H_B$ of each sector form a fundamental representation of a global group $SU(4)$ whose
 spontaneous symmetry breaking down to $SU(3)$ generates seven Goldstone bosons.
Six of them are eaten by $SU(2)_{LA}$ and $SU(2)_{LB}$ gauge bosons of each sector, and one is identified as the observed Higgs boson.

An important ingredient for the twin Higgs mechanism to work is the $Z_2$ twin symmetry under which each particle of one sector is interchanged with the corresponding particle of the other sector.
The role of the twin symmetry is to prevent the explicit breaking terms of $SU(4)$ symmetry from introducing a quadratic divergence of the Higgs boson.

However, the twin symmetry has to be broken either explicitly or spontaneously for various phenomenological reasons including the observed Higgs signal strength\,\cite{Barbieri:2005ri,Burdman:2014zta, Craig:2015pha}.
The $SU(4)$ breaking scale $f$ should be at least about three times larger than the SM Higgs vacuum expectation value (vev).
Introducing soft $Z_2$ symmetry breaking term $\epsilon f^2 |H_A|^2$ can easily provide a misalignment of twin Higgs vevs, but it requires fine tuning of parameters with an order of $v_{\rm SM}^2/f^2$ where $v_{\rm SM}\simeq 246\,\GeV$ is the SM Higgs vev.

\begin{figure}[t] 
	\begin{center}
		\includegraphics[width=0.4\textwidth]{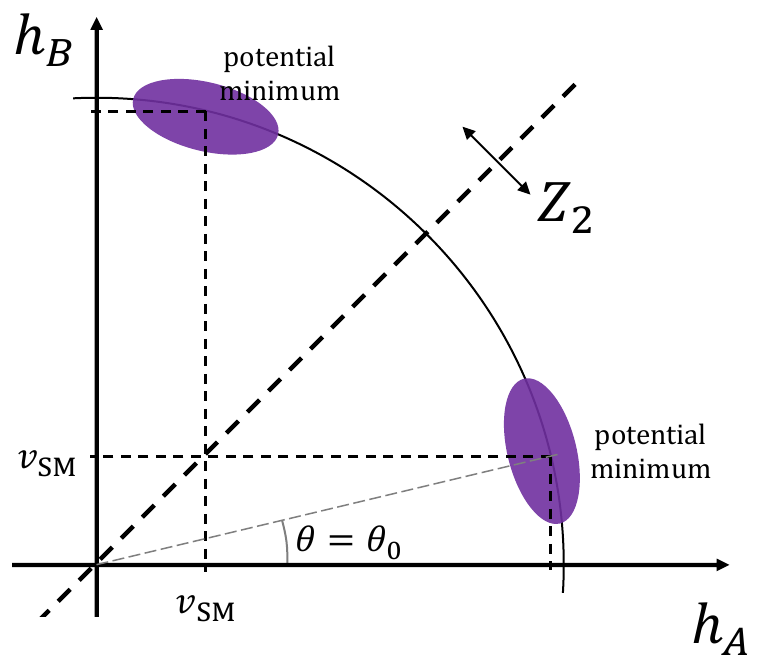} 
	\end{center}
	\caption{ Schematic picture of the potential for the spontaneous $Z_2$ breaking scenario. 
	There are two degenerate minima (purple) in $(h_1,\,h_2)$ field space. 
	Locations of minima are $Z_2$ symmetric.
	Vacuum chooses one of the minima before/during the inflation, and the sector with smaller vev becomes the SM.
	Circular line denotes the flat direction which corresponds to the Higgs boson degree.
	}
	\label{fig:scheme}
\end{figure}

In this paper, we focus on the possibility that the twin symmetry is exact but broken spontaneously.
Fig.\,\ref{fig:scheme} describes the desired situation that we consider in this paper.
In the twin Higgs field space $(h_A,\,h_B)$, the twin symmetry ($Z_2$) corresponds to the mirror symmetry along the diagonal dashed line.
There are two degenerate minima in the flat direction $h_A^2+h_B^2= f^2$ whose locations in the field space are symmetric under the $Z_2$ transformation.
Once scalar fields fall down to one of the minima, the sector with smaller Higgs vev becomes what we call the SM.
While there are several realizations in twin two Higgs doublet model\,\cite{Beauchesne:2015lva, Yu:2016bku} or singlet extended twin Higgs model\,\cite{Bishara:2018sgl},
 we focus on the realization in the minimal twin Higgs setup without additional scalar fields.
Similar idea is discussed in Refs.\,\cite{Hall:2018let, Dunsky:2019api} in the context of the strong CP problem.
We provide a more systematic approach to the construction of the effective potential.

Cosmological history of spontaneous twin symmetry breaking is strongly restricted by dark radiation constraints from the cosmic microwave background\,\cite{Aghanim:2018eyx}, and by the domain wall problem\,\cite{Zeldovich:1974uw}.
To avoid these problems, we assume that $Z_2$ symmetry is spontaneously broken before the end of the inflation, and
 reheaton decays mostly to the sector with smaller Higgs vev.
Possible realizations of such an asymmetric reheating can be seen in Ref.\,\cite{Craig:2016lyx, Chacko:2018vss}.
For preventing twin sector particles from being produced thermally, reheating temperature should be less than around the bottom quark mass when $f\sim 10\,v_{\rm SM}$.
Otherwise, twin sector particles can be produced through the bottom quark annihilation to the twin muon production process,
 and twin photons and twin neutrinos will finally contribute to the dark radiation\,\cite{Barbieri:2005ri,Craig:2016lyx, Chacko:2018vss, Barbieri:2016zxn,Chacko:2016hvu}.

\section{Minimal model with exact twin symmetry}

Explicit breaking of the global $SU(4)$ is necessary for the nonzero Higgs mass.
Without violating $Z_2$ symmetry, there are three sources of explicit $SU(4)$ breaking: quartic interaction, gauge interaction and Yukawa interaction.
To be more specific, let us consider an effective scalar potential, 
\bea
V(h_A,\,h_B)=\frac{\l}{4} \left( h_A^2+h_B^2-f^2\right)^2 +\Delta V(h_A,\,h_B),
\label{generalpotential}
\eea
where $h_A$ and $h_B$ are classical Higgs field of each sector.
Here, $\Delta V$ denotes terms breaking $SU(4)$ explicitly, but preserving $Z_2$.
A mixed quartic term $h_A^2h_B^2$ is the leading term to break $SU(4)$\footnote{$h_A^2 h_B^2$ is equivalent to $-(h_A^4+h_B^4)/2$ if we redefine $\l$ and $f$.}.
Gauge and Yukawa interactions will also contribute to $\Delta V$ through loops.

By replacing $h_A=f \cos \theta$ and $h_B=f \sin \theta$, we can obtain the potential along the flat direction for the case when $V\gg \Delta V$.
We denote it by $f^4 \hat V(\theta)$ in this paper.
Since $\hat V(\theta)$ is a periodic function with periodicity $\pi/2$ and symmetric under $Z_2:\,\theta\to \pi/2-\theta$, we can apply the Fourier expansion
\bea
\hat V(\theta) = \sum_n c_n \cos (4 n \theta)
\label{Fourier}
\eea
with coefficients $c_n$.

The leading order contribution $h_A^2 h_B^2$ corresponds to $(1-\cos 4\theta)/8$ whose extrema are $0$ ($h_B=0$), $\pi/4$ ($h_A=h_B$) and $\pi/2$ ($h_B=0$).
If its minima are at $0$ or $\pi/2$, $h_B$ or $h_A$ become zero and the electroweak symmetry breaking does not take place.
In order to obtain a proper misalignment, there should be nonzero $c_n$ contributions with $n\geq 2$ \footnote{
 If there are two Higgs doublets in each sector (i.e. $H_{1A}$, $H_{2A}$, $H_{1B}$, $H_{2B}$), a proper misalignment is possible by
 assigning different signs of mixed quartic couplings \cite{Beauchesne:2015lva}.
 }.

The simplest term to generate $c_{n\geq2}$ is the Coleman-Weinberg potential\,\cite{Coleman:1973jx} which is proportional to $\frac{1}{2}(h_A^4 \log h_A^2/\mu^2+h_B^4 \log h_B^2/\mu^2)$.
It leads to $c_1\propto (25-24\log 2)/96,~ c_2 \propto-1/240,~\cdots$ when we take $\mu=f$.
Since we want $c_2$ to have a sizable effect, a suppression of $c_1$ is required.
For this reason, we need a cancelation between contribution to $c_1$ from the $h_A^2 h_B^2$ term and the one from the Coleman-Weinberg potential.
This cancelation causes an unavoidable tuning of parameters in this scenario.
It will be shown that this cancelation is actually equivalent to the fine tuning of quadratic Higgs term in the infra-red (IR) theory.

In addition, the sign of $c_2$ should be positive for the spontaneous twin symmetry breaking.
If $c_2$ were negative, the minima could be only at $\theta = 0,\,\pi/4$ or $\pi/2$.
The positive sign of $c_2$ can be obtained if the beta function of Higgs self quartic coupling is negative.
It is noteworthy that the SM gauge and Yukawa interactions already provide the proper sign assignment.
Therefore, just an exact copy of the SM with a renormalizable Higgs portal potential \eqref{generalpotential} is enough to realize the spontaneous twin symmetry breaking.

The $SU(4)$ breaking effective potential $\Delta V$ can be parametrized by
\begin{equation}
\Delta V= \frac{\l_{\rm mix}}{4} h_A^2 h_B^2 + \frac{\beta}{4} \left( h_A^4 \log \frac{h_A}{\mu}+h_B^4 \log \frac{h_B}{\mu} \right)
\label{minimalV}
\end{equation}
where $\beta\equiv d\l / d \ln \mu$ is the renormalization group (RG) equation of $\l$.
Its SM value is 
\bea
\beta_{\rm SM} = \frac{1}{16\pi^2}\left( -6y_t^4+ \frac{9}{8}g_2^4+\frac{27}{200}g_1^4+\frac{9}{20}g_1^2g_2^2 \right),
\eea
where we ommit contributions coming from $\lambda_{\rm mix}$ because it will turn out to be negligible at $\mu=f$.
Note that $\lambda$ does not contribute to $\hat V(\theta)$ since $\lambda$ repects $SU(4)$ symmetry.
In Eq.\,\eqref{minimalV}, we redefined $\l_{\rm mix}(\mu=f)$ so that quartic operators coming from radiative contributions are absorbed, e.g. $\log y_t/2 -3/2$.

By taking $h_A=f \cos \theta$, $h_B=f \sin \theta$ and $\mu=f$, we obtain
\beq
\hat V = -\frac{\lambda_{\rm mix}}{32}\cos 4\theta + \frac{\beta}{4}\left(\cos^4\theta \log \cos \theta +\sin^4 \theta \log \sin\theta\right),
\label{Vhatminimal_exact}
\eeq
where we neglect the constant term for simplicity.
In terms of Fourier expansion, $\hat V$ can be written as
\beq
\hat V \simeq \frac{-12\l_{\rm mix}+\beta (25-24\log2)}{384} \cos 4\theta -\frac{\beta}{960} \cos 8\theta +\cdots.
\label{Vhatminimal}
\eeq

For the misalignment of vevs (i.e. $h_A\neq h_B \neq 0$), $\cos 4\theta$ term should be suppressed.
This suppression comes from the cancelation between $12\l_{\rm mix}$ and $\beta(25-24\log2)$.
In terms of $\kappa=\l_{\rm mix}/\beta$, the condition for the misalignment becomes
\bea
\beta<0 \text{~~and~~} \frac{1}{2}<\kappa< \frac{3-2\log2}{2}\simeq 0.81,
\eea
where the first condition $\beta<0$ is satisfied by the large top Yukawa interaction.
Here, we used $\hat V''(\pi/4)<0$ and $\hat V''(0)<0$ with $\hat V$ in Eq.\,\eqref{Vhatminimal_exact}.

\begin{figure}[t] 
\begin{center}
\includegraphics[width=0.47\textwidth]{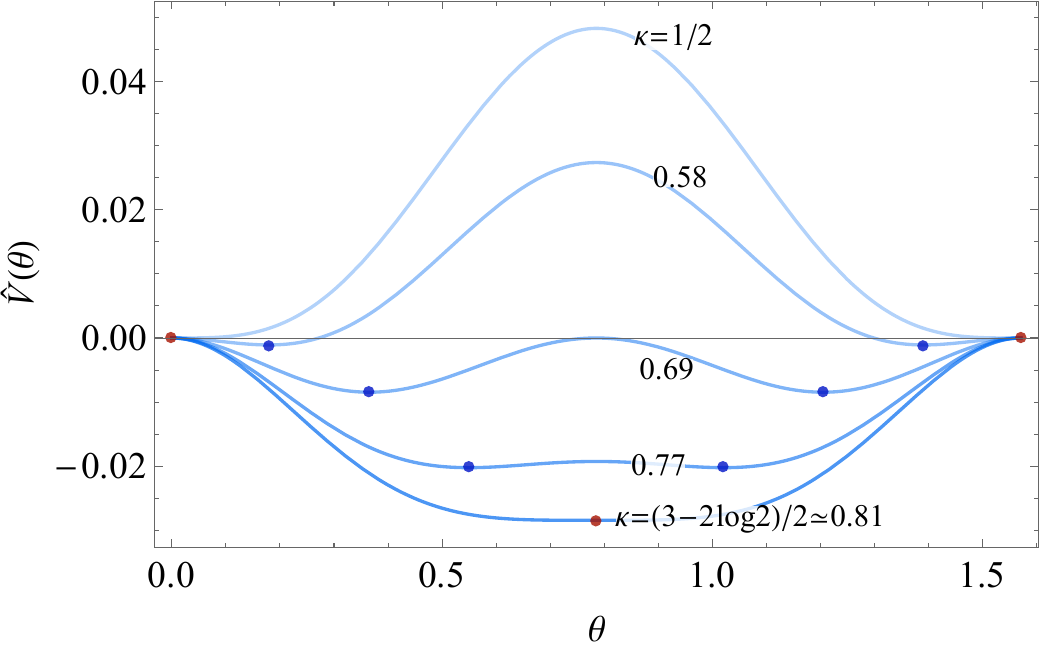} 
\end{center}
\caption{ Potential $\hat V$ along the flat direction is plotted with different $\kappa=\l_{\rm mix}/\beta$ values for a fixed $\beta=\beta_{\rm SM}$.  
If $\kappa<1/2$, potential minima are located at $\theta=0$ and $\pi/2$ which correspond to $v_{\rm SM}=0$.
For $\kappa>(3-2\log2)/2\simeq 0.81$, potential minimum is at $\theta=\pi/4$ where $Z_2$ symmetry is not boken spontaneously.}
\label{fig:Vhat}
\end{figure}

In Fig.\,\ref{fig:Vhat}, $\hat V(\theta)$ is described for different $\kappa$ values.
If $\kappa$ is too large, twin Higgs vevs become identical, i.e. $v_A=v_B$, and the twin symmetry is not broken spontaneously.
On the other hand, too small $\kappa$ leads one of twin Higgs vevs to zero, i.e. electroweak symmetry breaking does not occur.

\begin{figure}[t] 
\begin{center}
\includegraphics[width=0.45\textwidth]{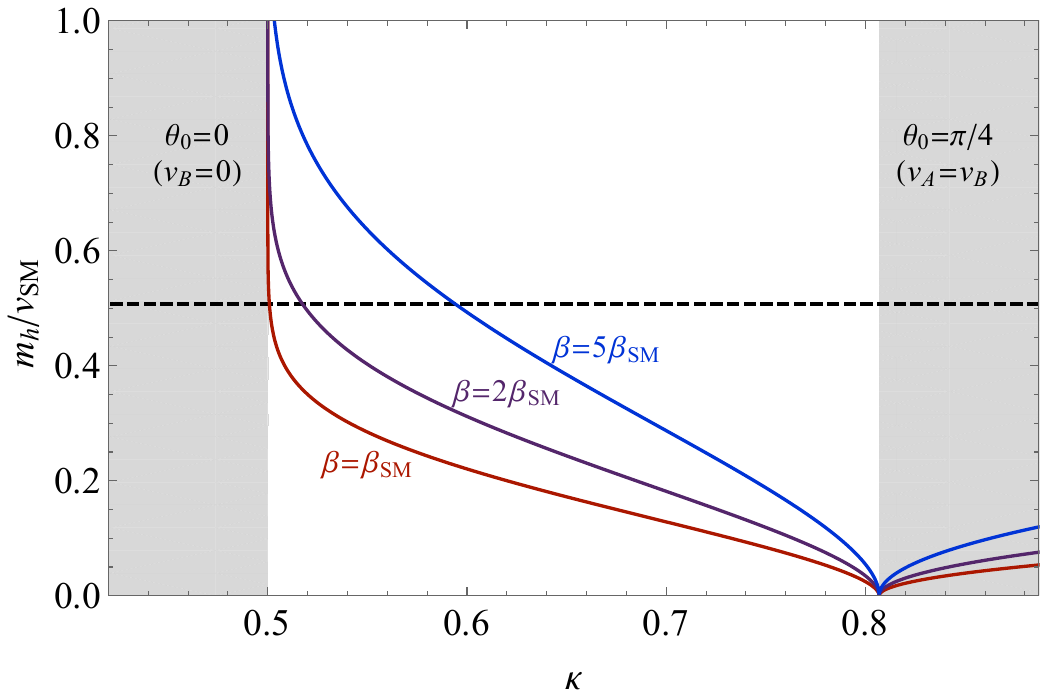} 
\end{center}
\caption{ $m_h/v_{\rm SM}$ is plotted as a function of $\kappa$ with different $\beta=\beta_{\rm SM}$ (red), $2\beta_{\rm SM}$ (purple) and $5\beta_{\rm SM}$ (blue).
Dashed line indicates the observed value, $m_h/v_{\rm SM}=125/246$.
For the minimal model (blue), $\kappa$ should be on the left corner from which one can read that a large fine tuning is required.
The situation is worse than this plot because $\beta_{\rm SM}$ is over-estimated as we take $\mu=M_Z$.
 }
\label{fig:massvevratio}
\end{figure}

In the minimal setup, $\beta=\beta_{\rm SM}$ and we have only two free parameters ($f$ and $\l_{\rm mix}$), and they are fixed by two observational constraints (Higgs vev $v_{\rm SM}$ and mass $m_h$).
If we denote $\theta_0$ as the minimum position of $\hat V$,
\bea
v_{\rm SM}&=&f \,{\rm min} (\sin \theta_0,\, \cos \theta_0),  \label{vh}\\
m_h^2&=& f^2\left.\frac{\partial^2 \hat{V}}{\partial \theta^2}\right|_{\theta=\theta_0}. 
\label{mh}
\eea
In Fig.\,\ref{fig:massvevratio}, $m_h/v_{\rm SM}$ is described as a function of $\kappa$ with fixed beta functions $\beta=\beta_{\rm SM}$ (red), $\beta=2\beta_{\rm SM}$ (blue) and $\beta=5\beta_{\rm SM}$ (blue).
For the minimal case ($\beta=\beta_{\rm SM}$), it is very difficult to obtain the observed value, $m_h/v_{\rm SM}\simeq 0.5$
 unless $\lambda_{\rm mix}$ stands at the edge of the allowed region.
In this plot, we estimated $\beta_{\rm SM}$ at $Z$ boson mass scale ($\mu=M_Z$), so more precise estimation will make the situation worse.
 
To investigate further, we expand the potential around $\theta=0$ with assuming $\kappa \simeq 1/2$, and we obtain
\beq
\hat V \simeq \frac{\beta}{4} \left( \kappa - \frac{1}{2} \right) \theta^2 +\beta \left(\frac{11}{48}-\frac{1}{3}\kappa \right) \theta^4 + \frac{\beta}{4} \theta^4 \log \theta
\label{Vhat_pol}
\eeq
where the first two terms are negative and the last logarithmic term is positive near $\theta \simeq 0$.
By multiplying $f^4$ and replacing $f \theta = h$, we can match Eq.\,\eqref{Vhat_pol} to the SM potential
\bea
f^4 \hat V \simeq -\frac{m^2}{2} h^2 + \frac{\lambda_h}{4} h^4 + \frac{\beta}{4}h^4\log \frac{h}{\mu},
\eea
where
\beq
m^2 = -\frac{\beta}{4}\left(\kappa-\frac{1}{2} \right)f^2, ~~~
\lambda_h (\mu=f) = \frac{1}{16}\beta. \label{matching}
\eeq
For the observed Higgs mass and vev, we need $m \simeq m_h/\sqrt{2}$ and $\l_h (\mu=M_Z) \simeq m_h^2/2v_{\rm SM}^2$.
Thus, we obtain
\bea
\kappa - \frac{1}{2} = \frac{2}{(-\beta)} \frac{m_h^2}{f^2}, 
\label{finetuning}
\eea
and $f$ is determined by the scale where $\l_h(\mu=f)= \beta/16 \lsim 0$ with boundary condition $\l_h (\mu=M_Z) \simeq m_h^2/2v_{\rm SM}^2$.
The prediction of the minimal model is $f \simeq 10^{10}\,\GeV$\,\cite{Buttazzo:2013uya}.
The metastability of the Higgs boson at the IR theory (SM) can be interpreted as a consequence of the spontaneous twin symmetry breaking.

However, the Eq.\,\eqref{finetuning} tells that $\l_{\rm mix}$ needs to be very close to $\beta/2$ (remind that $\kappa = \lambda_{\rm mix}/\beta$).
Since there is no reason for this relation, it should be regarded as a tuning.\footnote{
The tuning of quartic coupling $\l_{\rm mix}$ in ultra-violet (UV) theory represents the tuning of Higgs quadratic coupling in the IR theory as can be seen in Eq.\,\eqref{matching}.}
The order of tuning is alleviated by the factor of $\beta$ compared to other twin Higgs scenarios, but is basically ${\cal O}(m_h^2/f^2)$.
For the theory to be natural, the scale $f$ should not be very far away from the weak scale.
In the next section, we will discuss one example to make $f$ low.

\section{Vectorlike leptons}

A possible way to alleviate tuning is introducing new Yukawa interactions.
Additional Yukawa interactions can give negative contributions to $\beta$, and make $f$ smaller.
It can also be seen in Fig.\,\ref{fig:massvevratio} that if $\beta$ is larger than the SM value (purple and blue curves), the slope at $m_h/v_{\rm SM}$ could be small, so the tuning of $\lambda_{\rm mix}/\beta$ can be milder.

As an example, we consider a family of vectorlike leptons (VLL) in each sector:
 lepton doublets $L_{Li}$, $\bar L_{Li}$, charged lepton singlets $E_{Ri}$, $\bar E_{Ri}$, neutral lepton singlets $N_{Ri}$ and $\bar N_{Ri}$ for $i=A,\,B$.
Their interaction Lagrangian can be written as
\bea
\hspace{-1cm}
&&\hspace{-0.2cm} {\cal L}_{\rm VLL}= - M_L \bar L_{Li} L_{Ri} -M_E \bar E_{Li} E_{Ri} -M_N \bar N_{Li} N_{Ri}
 \\
&&\hspace{-0.2cm}- y_E \bar L_{Li} E_{Ri} H_i  - \bar y_E \bar E_{Li} L_{Ri} H_i^\dagger 
- y_N \bar L_{Li} N_{Ri} H_i^\dagger- \bar y_N \bar N_{Li} L_{Ri} H_i, \nn
\eea
where the summation of $i=A,\,B$ is ommitted in the expression.
 Although there are several implications of VLL for the case when they couple to the SM leptons\,\cite{Fujikawa:1994we,Kannike:2011ng,Dermisek:2013gta,Poh:2017tfo,DEramo:2007anh,Cohen:2011ec,Restrepo:2015ura,Calibbi:2015nha,Bhattacharya:2015qpa,Bhattacharya:2017sml}, we do not discuss them in this paper because their coupling to the SM leptons should be small anyway, so their contributions to the effective potential are negligible.

Mass matrix of charged and neutral VLL are given by
\beq
{\cal M}_{Ei}=
\left( \begin{array}{cc}
M_L & \frac{y_E v_i}{\sqrt{2}}\\
\frac{\bar y_E v_i}{\sqrt{2}} & M_E
\end{array} \right),~~
{\cal M}_{Ni}=
\left( \begin{array}{cc}
M_L & \frac{y_N v_i}{\sqrt{2}}\\
\frac{\bar y_N v_i}{\sqrt{2}} & M_N
\end{array} \right).
\label{mass}
\eeq
For simplicity, we assume that $M_L=M_E=M_N$ and $y_L \equiv y_E= y_N$ and $\bar y_E = \bar y_N=0$\footnote{This parameter choice is only for the simplicity. Later of this section, we estimate phenomenological constraints such as Peskin-Takeuchi parameters or Higgs to diphoton signal rate in a general parameter space.}.
The eigenvalues of ${\cal M} {\cal M}^\dagger$ become
\bea
M_{Li \pm}=M_L^2 + \frac{v_i^2y_L^2}{4} \pm \frac{1}{4}\sqrt{8M_L^2 y_L^2 v_i^2+y_L^4 v_i^4},
\eea
for $i=A,\,B$.

The effective potential from VLL in each sector is given by
\beq
-8\times \frac{1}{64\pi^2} \left( M_{Li+}^4 \log \frac{M_{Li+}^2}{\mu^2}+ M_{Li-}^4 \log \frac{M_{Li-}^2}{\mu^2} \right)
\label{effV_VLL}
\eeq
which can provide large enough $c_2$ when $M_L \lsim f$.
If $M_L \gsim f$, $c_2$ become suppressed by $(f/M_L)^4$, and Eq.\,\eqref{effV_VLL} can give only the threshold correction to $\l_{\rm mix}$.
Thus, $M_L$ should be close to the $SU(4)$ breaking scale.

Collider signatures of VLL are highly sensitive on the mixing with SM leptons\,\cite{AguilarSaavedra:2009ik,Redi:2013pga,Falkowski:2013jya,Holdom:2014rsa,Kumar:2015tna,Bertuzzo:2017wam,Kling:2018wct}.
The mass limit for charged leptons from the Large Electron-Positron collider (LEP) is $100.8\,\GeV$  when the charged lepton mostly decays to $W\nu$.
For neutral leptons, the mass limit from LEP is $101.3\,\GeV$ when they decay to $We$\,\cite{Achard:2001qw, Tanabashi:2018oca}.
At Large Hadron Collider (LHC), the most relevant search is Refs.\,\cite{Sirunyan:2018mtv} which provides constraints on the CKM matrix elemant $|V_{eN}|$ and $|V_{\mu N}|$ in the mass range from GeV to TeV.
At this moment, LHC constraints are comparable to LEP constraints\,\cite{Sirunyan:2018mtv}, but there will be much improvement in the future.

 \begin{figure}[t] 
\begin{center}
\includegraphics[width=0.47\textwidth]{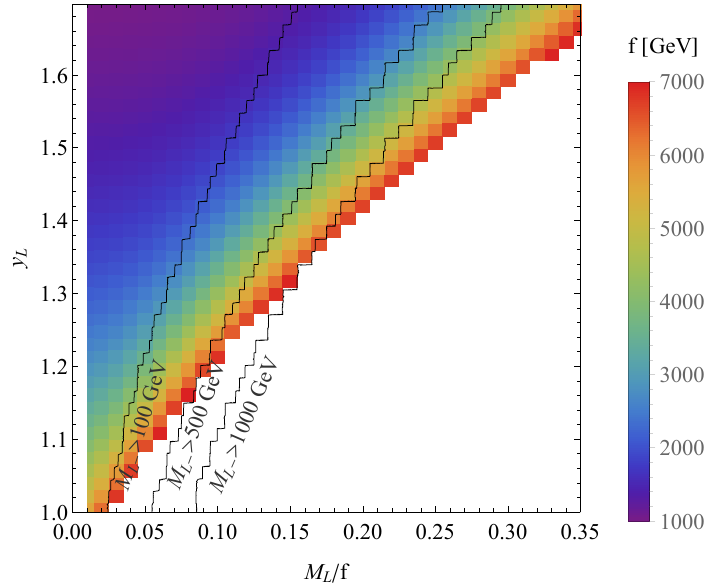} 
\end{center}
\caption{ $SU(4)$ breaking scale $f$ is determined in the $(M_L/f,\, y_L)$ parameter space. Here, we choose $y_E=y_N=y_L$, $\bar{y}_E=\bar{y}_N=0$ and $M_L=M_E=M_N$. Colors represent the scale of $f$. Black solid lines correspond to lightest lepton mass $100\,\GeV$, $500\,\GeV$ and $1\,\TeV$.}
\label{fig:final}
\end{figure}

Fig.\,\ref{fig:final} shows the $SU(4)$ breaking scale $f$ by colors in the parameter space of $M_L/f$ and $y_L$.
Black solid lines correspond to lightest lepton mass $100\,\GeV$, $500\,\GeV$ and $1\,\TeV$.
We restrict Yukawa coupling to be smaller than $1.7$ because the Landau pole arises below $100\,\TeV$ if it becomes larger.
The smallest $f$ within $m_{L-}>100\,\GeV$, $500\,\GeV$ and $1\,\TeV$ are $f\gsim 1.3\,\TeV$, $2.7\,\TeV$ and $3.9\,\TeV$, respectively.

Another advantage of VLL comes from changing RG running of top Yukawa coupling.
In the minimal model, top Yukawa coupling rapidly drops because of large $SU(3)_c$ gauge coupling.
If there are additional Yukawa interactions, they compensate negative contributions of gauge coupling and make the pseudo-IR fixed point smaller.
Consequently, top Yukawa coupling can maintain its strength until $\mu\sim f$.
We neglect this effect in Fig.\,\ref{fig:final}, so $f$ can be slightly smaller value in more precise calculations.

\begin{figure}[t] 
\begin{center}
\includegraphics[width=0.43\textwidth]{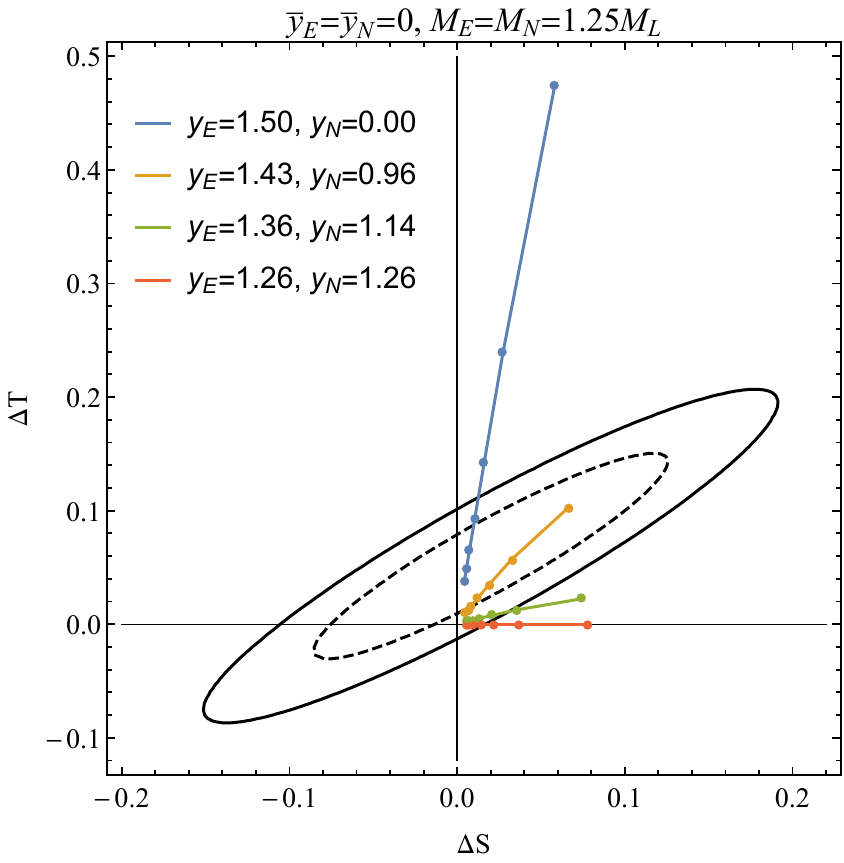}
\includegraphics[width=0.43\textwidth]{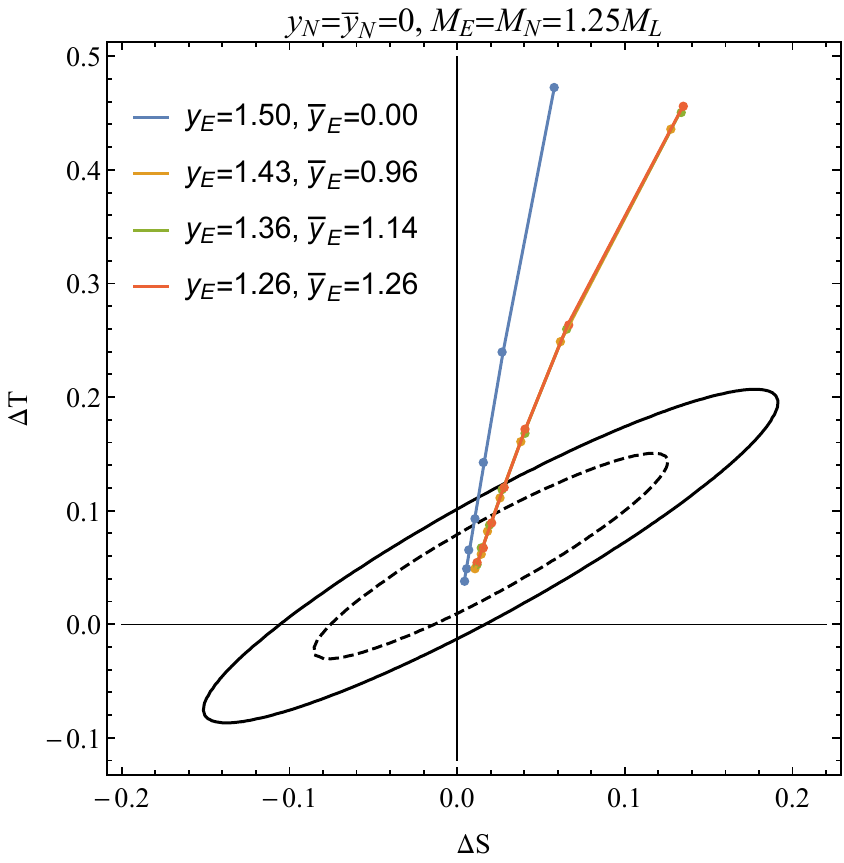}
\end{center}
\caption{Peskin Takeuchi parameters $\Delta S$ and $\Delta T$ are described with parameter choices $\bar y_E=\bar y_N=0$ (upper plot) and $y_N=\bar{y}_N$ (lower plot). 
In both cases, we take $M_E=M_N=1.25M_L$. Black dashed (solid) lines correspond to the current bound on $\Delta S$ and $\Delta T$ in 68\,\% (95\,\%)  confidence level \cite{Tanabashi:2018oca}.
Points represent the lightest vectorlike lepton masses $M_{L-}=100,$ 200, 300 $\cdots$ 700 GeV from the right to the left.}
\label{ST}
\end{figure}

VLLs with large Yukawa couplings can modify electroweak precision parameters and Higgs to diphoton signal strength through the loop processes.
We estimate constraints coming from the electroweak precision by using Peskin-Takeuchi parameters $\Delta S$ and $\Delta T$ \,\cite{Peskin:1990zt, Tanabashi:2018oca}.
Detailed calculations are summarized in the Appendix.\,\ref{app:ST}.
Numerical results with various parameter choices are described in Fig.\,\ref{ST}.
Here, we fix $y_{\rm eff}\equiv \left(\frac{1}{2} \sum_i |y_i|^4\right)^{1/4}=1.26$, which should correspond to $y_L$ in Fig.\,\ref{fig:final}.

Since $\Delta T$ represents the strength of custodial symmetry breaking in the new physics, $y_E=y_N$ and $\bar{y}_E=\bar{y}_N$ leads to $\Delta T=0$.
Without tuning of Yukawa couplings, we conclude that $M_{L-}\gsim 400$ GeV is safe for large Yukawa couplings $y_{\rm eff}=1.26$.
For even larger Yukawa couplings, naive estimation would be $M_{L-}\gsim y_{\rm eff}^2\times 350\,{\rm GeV} $ since $\Delta T\propto y^4$ in terms of Yukawa coupling differences among charged leptons and neutral leptons.

\begin{figure}[t] 
\begin{center}
\includegraphics[width=0.47\textwidth]{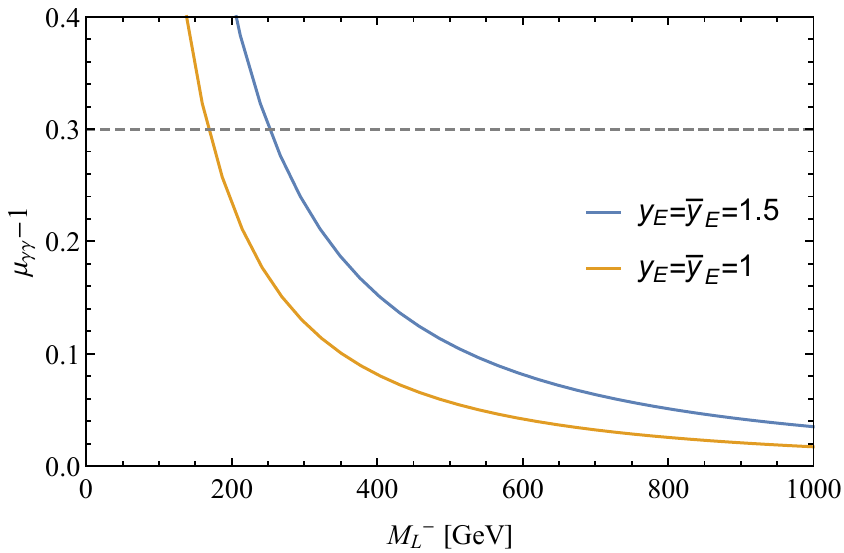} 
\includegraphics[width=0.47\textwidth]{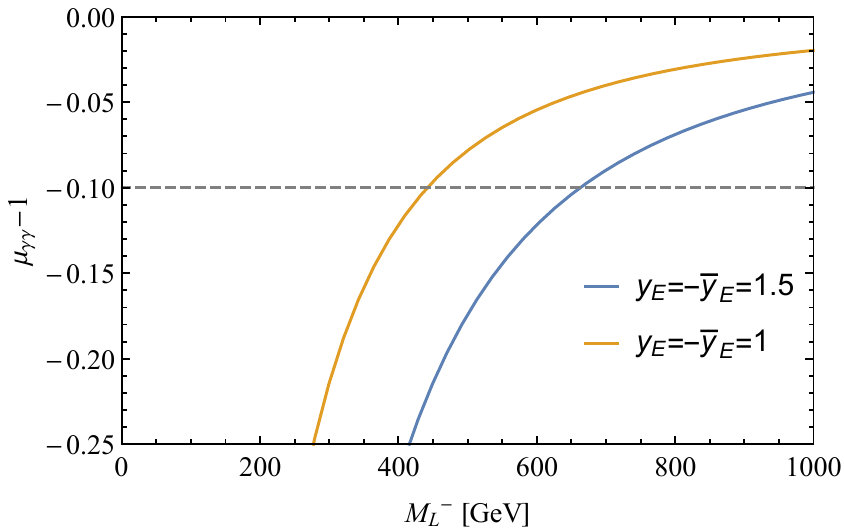} 
\end{center}
\caption{The signal strength $\mu_{\gamma\gamma}$ is drawn as a function of the lightest vectorlike lepton mass.
From bottom to top, we take $y_E=\bar y_E$ (upper plot) and $y_E=-\bar y_E$ (lower plot) to be 1 and 1.5.
In each case, we take $M_E=M_N=1.25M_L$. 
Horizontal dashed line corresponds to the current 2$\sigma$ bound obtained in Ref.\,\cite{Cheung:2018ave}.}
\label{diphoton}
\end{figure}

If the charged lepton Yukawa coupling is large, Higgs to diphoton signal can be modified significantly.
Here, we estimate Higgs to diphoton signal strength which can be obtained by
\bea
\mu_{\gamma\gamma}=\frac{\Gamma_{h\to \gamma \gamma}}{\Gamma_{h\to \gamma \gamma}^{\rm SM}},
\eea
where the denominator (numerator) is the decay rate of Higgs to diphoton in the SM (the model with VLLs).
Formulas used in the numerical calculation can be found in the Appendix.\,\ref{app:diphoton}.
In this paper, we estimate both of $\Gamma_{h\to \gamma \gamma}^{\rm SM}$ and $\Gamma_{h\to \gamma \gamma}$ in one loop order.
The numerical results of $\mu_{\gamma\gamma}$ are depicted as functions of the lightest VLL mass $M_{L-}$ in Fig.\,\ref{diphoton}.
In this plot, we simply take $y_E=\bar y_E$.
Horizontal lines correspond to the current 2$\sigma$ bound obtained in Ref.\,\cite{Cheung:2018ave} which yields $\mu_{\gamma \gamma}=1.1 \pm 0.10$ in $1\sigma$ error.
From the figure, we obtain a bound on the lightest VLL mass $M_{L-}\gsim 150$ - $250$ GeV for the Yukawa coupling $y_E=\bar y_E=$ 1 - 1.5 with the same sign.
If $y_E \bar y_E<0$, the corresponding bound becomes $M_{L-}\gsim 450$ - $650$ GeV.

If there are mixings between SM leptons and VLLs, there can be richer phenomenological implications such as lepton flavor violations or anomalous magnetic momentum of muon.
However, the mixing should be small for the lepton flavor violations and thus contributions to the effective potential are negligible\,\cite{Poh:2017tfo}. The smallness of the mixing could be understood in terms of symmetry argument, i.e. technical naturalness. For example, one can assign a charge of global symmetry in the the VLL sector.


\section{Conclusion}

We have discussed the possibility of spontaneous twin symmetry breaking scenario.
For the misalignment of nonzero twin Higgs vevs, there should be $\cos 8\theta$ term in $\hat V(\theta)$ which can come from the Coleman-Weinberg potential.
In addition, we need a cancellation of ${\cal O}(v^2/\beta f^2)$ between $\l_{\rm mix}$ and $\beta/2$.

The $SU(4)$ breaking scale $f$ is determined by the scale where the Higgs self quartic coupling flips its sign in the IR theory.
Since, in the SM, this flipping occurs around ${\cal O}(10^{10}\,\GeV)$, the minimal setup is not natural.
For obtaining smaller $f$, we introduced twin VLLs with large Yukawa couplings and obtained $f$ as small as $2.7\,\TeV$ when they are safe from the current bounds on VLLs.

Constraints on VLLs are estimated from the electroweak precision parameters $\Delta S$ and $\Delta T$ and Higgs to diphoton signal strength $\mu_{\gamma\gamma}$.
With strong enough Yukawa couplings, the lightest VLL mass should be larger than around $400$ GeV.
Potentially, VLLs are testable at the LHC or future lepton colliders through Higgs measurement and direct production depending on the mixing with SM leptons.
Their signatures below the TeV scale would lend credence to this scenario.

Although we do not specify the inflation sector, we have assumed that reheaton decays mostly to the sector with smaller Higgs vev in order to avoid cosmological problems.
The reheating temperature of the SM sector should be less than around bottom quark mass for preventing thermal production of twin sector particles.
An interesting possibility is that cogenesis of baryon asymmetry and asymmetric dark matter could occur during the reheating process.
We leave detailed studies about cosmological history and the inflation sector as future work.

\noindent{\bf Acknowledge}
THJ is grateful to Chang Sub Shin, Kyu Jung Bae and Dongjin Chway for useful discussions.
This work was supported by IBS under the project code, IBS-R018-D1.

\begin{appendix}
\section{Electroweak Precision Parameters}  \label{app:ST}
Yukawa couplings of VLLs to the Higgs field induce a mixing between left handed leptons and right handed leptons.
Let us denote $U_{EL}$ and $U_{ER}$ by the mixing matrix such that $U_{EL} {\cal M}_E U_{ER}^\dagger={\rm diag}(M_{E-}, M_{E+})$, and $U_{NL} {\cal M}_N U_{NR}^\dagger={\rm diag}(M_{N-}, M_{N+})$.
VLL interactions with the $Z$ boson are given by
\bea
g_{EL, R}^Z&=&\frac{g}{cw} U_{EL, R}
\left( \begin{array}{cc}
-\frac{1}{2}+sw^2 & 0 \\
0 & sw^2
\end{array} \right) U_{EL, R}^\dagger ,   \label{gZE}  \\
g_{NL, R}^Z&=&\frac{g}{cw} U_{NL, R} ,
\left( \begin{array}{cc}
\frac{1}{2} & 0\\
0 & 0
\end{array} \right) U_{NL, R}^\dagger  ,  \label{gZN}  
\eea
where $sw$ and $cw$ are sin and cos of the Weinberg angle, and $g$ is the $SU(2)_W$ gauge coupling constant.
For the final result to be consistent with the renormalization condition, we take $cw=M_W/M_Z$.
With the $W$ boson, we have 
\bea
g_{L}^{W^+}=\frac{g}{\sqrt{2}} U_{NL, R}
\left( \begin{array}{cc}
1 & 0\\
0 & 0
\end{array} \right) U_{EL, R}^\dagger.  \label{gW}  
\eea
Note that the photon interaction remains $Q_{EM} g\,sw$, and diagonal, i.e. vectorlike, in the basis of mass eigenstates.

The electroweak precision parameter $S$ and $T$ are defined as
\bea
\alpha S&=&\frac{4 sw^2 cw^2}{M_Z^2}  \Big( \Pi_{ZZ}(M_Z^2)-\Pi_{ZZ}(0)\\ 
&&~ -\frac{cw^2-sw^2}{sw\, cw} \Pi_{Z\gamma}(M_Z^2)-\Pi_{\gamma\gamma}(M_Z^2) \Big) ,\\
\alpha T&=& \frac{\Pi_{WW}(0)}{M_W^2}-\frac{\Pi_{ZZ}(0)}{M_Z^2},
\eea
where $\Pi_{AB}(q^2)$ is the self energy diagram of external $A$ boson and $B$ boson, $\alpha$ is the fine structure constant, i.e. $\alpha = g^2 sw^2/4\pi$.

The finite piece of the self energy diagram in the $\overline{\rm MS}$ scheme is given by
\bea
\Pi^{LL}(q^2)&=&-\frac{4}{(4\pi)^2} \Big[-q^2 b_2(m_1^2,m_2^2)  \\
&&+\frac{1}{2}\left(m_2^2 b_1(m_1^2,m_2^2)+m_1^2 b_1(m_1^2,m_2^2)\right) \Big] ,\nn \\
\Pi^{LR}(q^2)&=&\frac{2}{(4\pi)^2} \big[ m_1 m_2 b_0(m_1^2, m_2^2) \big],
\eea
where the loop functions can be found in Ref.\,\cite{Peskin:1995ev},
\bea
b_0(m_1^2,m_2^2)&=&\int_0^1 dx \log\left(\Delta(m_1^2, m_2^2, q^2)/\mu^2\right) ,\\
b_1(m_1^2,m_2^2)&=&\int_0^1 dx x \log\left(\Delta(m_1^2, m_2^2, q^2)/\mu^2\right) ,\\
b_2(m_1^2,m_2^2)&=&\int_0^1 dx x(1-x) \log\left(\Delta(m_1^2, m_2^2, q^2)/\mu^2\right), \nn\\
\eea
with $\Delta(m_1^2,m_2^2,q^2)= x m_2^2+(1-x)m_1^2-x(1-x)q^2$.
Here, $LL$ or $LR$ denotes how the projection operator $P_L$ and $P_R$ are inserted in the left and right vertices of the diagram.
Since we have $\Pi^{LL}=\Pi^{RR}$ and $\Pi^{LR}=\Pi^{RL}$, one should combine up $\Pi^{LL}$ and $\Pi^{LR}$ with proper coefficients given by Eq.\,(\ref{gZE} - \ref{gW}).
If the couplings are correctly assigned, the RG scale $\mu$ dependence should be canceled out in $S$ and $T$ parameters.

\section{Higgs to Diphoton Decay Rate} \label{app:diphoton}
With a scalar loop function $f$ given by 
\bea
f(x)=
\left\{ \begin{array}{ll}
\arcsin^2 \sqrt{x}, & x\leq 1 \\
-\frac{1}{4}\left( \log\frac{1+\sqrt{1-x^{-1}}}{1-\sqrt{1-x^{-1}}} - i\pi \right)^2, & x>1
\end{array} \right. ,
\eea
we have the $h\to \gamma\gamma$ triangle diagram with a fermion loop $A_{1/2}$ and a boson loop $A_{1}$ given by
\bea
A_{1/2}(x)&=&\frac{2}{x^2}(x+(x-1)f(x)) ,\\
A_1(x)&=&-\frac{1}{x^2} (2x^2+3x+3(2x-1) f(x)),
\eea
where $x=m_h^2/4M_i^2$ with the loop particle mass $M_i$ and the Higgs mass $m_h=125$ GeV\,\cite{Djouadi:2005gi}.

The decay rate of $h\to \gamma\gamma$ in the SM is given by
\bea
\Gamma^{\rm SM}(h\to \gamma\gamma)= \frac{G_\mu \alpha^2 m_h^3}{128 \sqrt{2}\pi^3}
\left| \sum_f N_c Q_f^2 A_{1/2}(x_f) + A_1 (x_W) \right|^2 , \nn\\
\eea
where $G_\mu$ is the Fermi constant.
If there are vectorlike leptons, we have 
\bea
\Gamma^{\rm SM}(h\to \gamma\gamma)= \frac{G_\mu \alpha^2 m_h^3}{128 \sqrt{2}\pi^3}
\left| A_{\rm SM}+  \sum_{i\in {\rm VLL}} \frac{Y_{ii} v_{\rm SM}}{\sqrt{2}M_{Ei}} A_{1/2}(x_{Ei}) \right|^2 , \nn\\
\eea
where $A_{\rm SM}=\sum_f N_c Q_f^2 A_{1/2}(x_f) + A_1 (x_W)$ and $M_i$ is the VLL mass eigenvalue with $i=\pm$, and the Yukawa matrix $Y$ is given by
\bea
Y=U_{EL} \left( \begin{array}{cc}
0 & y_E \\
\bar{y}_E & 0
\end{array} \right) U_{ER}^\dagger.
\eea

\end{appendix}

\end{document}